\documentclass[apjl]{emulateapj}
\pdfoutput=1
\usepackage[varg]{txfonts}
\setkeys{Gin}{width=0.82\linewidth, height=0.8\textheight, keepaspectratio=true}

\newcommand\Hmol{\ensuremath{\mathrm{H_2}}}
\newcommand\pcc{\ensuremath{\mathrm{cm^{-3}}}}
\newcommand\kms{\ensuremath{\mathrm{km\ s^{-1}}}}
\newcommand\hii{\ion{H}{2}}
\newcommand\lamad{\ensuremath{\lambda_\mathrm{ad}}}

\begin{document}

\title{Merged ionization/dissociation fronts in planetary nebulae\altaffilmark{1}}
\author{%
  William J. Henney,\altaffilmark{2}
  R. J. R. Williams,\altaffilmark{3}
  Gary J. Ferland,\altaffilmark{4}
  Gargi Shaw,\altaffilmark{4}
  \& C. R. O'Dell\altaffilmark{5}
}
\altaffiltext{1}{Contains material \textcopyright{} Crown Copyright 2007/MOD.}

\altaffiltext{2}{Centro de Radioastronom\'{\i}a y Astrof\'{\i}sica,
  Univer\-si\-dad Na\-cio\-nal Aut\'o\-no\-ma de M\'exi\-co, Morelia,
  M\'exico} %

\altaffiltext{3}
{AWE, Aldermaston, RG7 4PR, UK}

\altaffiltext{4}{Department of Physics and Astronomy, University of
  Kentucky, Lexington, KY 40506}

\altaffiltext{5}{Department of Physics and Astronomy, Vanderbilt
  University, Box \mbox{1807-B}, Nashville, TN 37235}

\shorttitle{Merged ionization/dissociation fronts}
\shortauthors{Henney et al.}

\email{w.henney@astrosmo.unam.mx, rjrwilliams@googlemail.com,
  gary@uky.edu, gargishaw@gmail.com, cr.odell@vanderbilt.edu}

\begin{abstract}
  The hydrogen ionization and dissociation front around an ultraviolet
  radiation source should merge when the ratio of ionizing photon flux
  to gas density is sufficiently low and the spectrum is sufficiently
  hard. This regime is particularly relevant to the molecular knots
  that are commonly found in evolved planetary nebulae, such as the
  Helix Nebula, where traditional models of photodissociation regions
  have proved unable to explain the high observed luminosity in H$_2$
  lines. In this paper we present results for the structure and
  steady-state dynamics of such advection-dominated merged fronts,
  calculated using the Cloudy plasma/molecular physics code. We find
  that the principal destruction processes for H$_2$ are
  photoionization by extreme ultraviolet radiation and charge exchange
  reactions with protons, both of which form H$_2{}^+$, which rapidly
  combines with free electrons to undergo dissociative
  recombination. Advection moves the dissociation front to lower
  column densities than in the static case, which vastly increases the
  heating in the partially molecular gas due to photoionization of
  He$^0$, H$_2$, and H$^0$. This causes a significant fraction of the
  incident bolometric flux to be re-radiated as thermally excited
  infrared \Hmol{} lines, with the lower excitation pure rotational
  lines arising in $1000$~K gas and higher excitation \Hmol{} lines
  arising in $2000$~K gas, as is required to explain the H$_2$
  spectrum of the Helix cometary knots.
\end{abstract}


\defcitealias{2007AJ....133.2343O}{OHF07}
\defcitealias{2006ApJ...652..426H}{H06}

\section{Introduction}
\label{sec:introduction}

The ultraviolet photons from hot stars, such as main-sequence OB stars
or the central stars of planetary nebula (CSPN) will dissociate and
ionize the surrounding circumstellar and interstellar gas. In the
classical picture \citep[e.g.,][]{1997ARA&A..35..179H}, the extreme
ultraviolet (EUV) photons with energies $h\nu > 13.6~\mathrm{eV}$
photoionize the hydrogen gas, forming an \hii{} region bounded by a
relatively sharp ionization front (IF), while the far ultraviolet
(FUV) photons with energies $6~\mathrm{eV} < h\nu < 13.6~\mathrm{eV}$
penetrate the IF to form a neutral photodissociation region (PDR) that
surrounds the \hii{} region. The dissociation of \Hmol{} in such a PDR
is dominated by a two-step radiative process
\citep{1967ApJ...149L..29S, 2000A&AS..141..297A}, in which absorption
of an FUV photon leaves the \Hmol{} molecule in an electronically
excited state, from which it has a certain probability ($\simeq 10$\%)
of decaying to the vibrational continuum of the ground electronic
state.

However, as shown by \citet{1996ApJ...458..222B}, a classical extended
PDR cannot exist if the FUV flux is sufficiently weak compared with
the EUV flux at the IF, rather the IF and dissociation front (DF)
should merge.  \citeauthor{1996ApJ...458..222B} considered the case of
\hii{} regions around OB stars and found that this regime was most
relevant for cases in which the dust optical depth through the ionized
gas is low, which corresponds to a low ionization parameter (the
ionization parameter is a dimensionless number equal to the ratio of
the number density of ionizing photons to the number density of
hydrogen nuclei). To date, no models have been calculated of the
structure and emission properties of such fronts, which are also
expected to show strong deviations from static chemical and ionization
equilibrium.

In this paper, we calculate in detail the internal dynamics and
chemistry of merged ionization/dissociation fronts (IF/DFs),
concentrating on the particular case of compact photoevaporating
molecular knots in evolved planetary nebulae (PNe), such as are seen
in the Helix nebula (\citealp{1999ApJ...522..387Y,
  2005AJ....130.1784M, 2006ApJ...652..426H};
\citealp[OHF07]{2007AJ....133.2343O}). The stellar spectrum from the
hot central star of a PN is harder than that from an OB star, leading
to an EUV luminosity that is much greater than the FUV luminosity. In
addition, the ionization parameter of the knots is much lower than is
typically seen in \hii{} regions, resulting in very little attenuation
of the EUV flux by recombinations in the ionized gas. Both these
factors place the knots firmly in the merged IF/DF regime. The most
comprehensive existing theoretical study of PDRs in planetary nebulae
is that of \citet{1998A&A...337..517N}, who present detailed modeling
of the time-dependent evolution of an expanding circumstellar shell as
the luminosity and spectrum of the CSPN evolves, finding that soft
X-rays can be important in exciting the molecular gas at late
times. However, \citetalias{2007AJ....133.2343O} showed that
this is not the case in the Helix since it is only in the EUV band
that the CSPN luminosity is sufficient to excite the knot PDRs.
\citeauthor{1998A&A...337..517N} used an analytic treatment of the
\hii{} region, which is assumed to absorb all the EUV radiation, and
were thus unable to model the case of a merged IF/DF.

\section{Models}
\label{sec:models}

\begin{table}[t]
  \newcommand\C[1]{\multicolumn{1}{c}{#1}}
  \newcommand\CC[1]{\multicolumn{6}{l}{#1}}
  \centering
  \setlength\tabcolsep{2\tabcolsep}
  \renewcommand\arraystretch{1.1}
  \caption{Model input parameters}
  \label{tab:input}
  \smallskip
  \vspace*{-\smallskipamount}
  \begin{tabular}{rrrrrr}
    \hline
    \CC{Stellar parameters}\\[\smallskipamount]
    \CC{$L = 120~L_\odot$ \quad
      $T_\mathrm{eff} = 130,000~\mathrm{K}$} \\
    \CC{$Q_\mathrm{H} = 7.57 \times 10^{45}~\mathrm{s^{-1}}$ \quad
      $Q_\mathrm{FUV} = 1.33 \times 10^{45}~\mathrm{s^{-1}}$} \\[\smallskipamount]
    \hline
    \raisebox{\medskipamount}{\phantom{A}}& \C{$D$} & \C{$u_0$}  &  \C{$n_0$} 
    & \C{$F_0$}  &  \\
    \C{Model}   & \C{(pc)} & \C{(\kms)} &  \C{(\pcc)}
    & \C{($\mathrm{cm^{-2}\ s^{-1}}$)} & \C{$\lambda_\mathrm{ad}$} \\
      \hline
      A10  & 0.137 & 10 & 3162. &  $3.36 \times 10^{9}$ & 0.94 \\
      A06  & 0.137 &  6 & 3162. &  $3.36 \times 10^{9}$ & 0.56 \\
      A03  & 0.137 &  3 & 3162. &  $3.36 \times 10^{9}$ & 0.28 \\
      A01  & 0.137 &  1 & 3162. &  $3.36 \times 10^{9}$ & 0.09 \\
      A00  & 0.137 &  0 & 3162. &  $3.36 \times 10^{9}$ & 0.00 \\
      AA10 & 0.137 & 10 & 1000. &  $3.36 \times 10^{9}$ & 0.30 \\
      B10  & 0.244 & 10 & 1000. &  $1.06 \times 10^{9}$ & 0.94 \\
      B06  & 0.244 &  6 & 1000. &  $1.06 \times 10^{9}$ & 0.57 \\
      B00  & 0.244 &  0 & 1000. &  $1.06 \times 10^{9}$ & 0.00 \\
      C10  & 0.433 & 10 &  316. &  $3.36 \times 10^{8}$ & 0.94 \\
      C06  & 0.433 &  6 &  316. &  $3.36 \times 10^{8}$ & 0.56 \\
      C00  & 0.433 &  0 &  316. &  $3.36 \times 10^{8}$ & 0.00 \\ \hline
  \end{tabular}
\end{table}

In order to investigate the structure of advective IF/DFs in PNe, we
have calculated steady-state, plane-parallel slab models using the
Cloudy plasma/molecular physics code
\citep{1998PASP..110..761F}. Details of the computational scheme used
to treat the steady-state dynamics are given in
\citet{2005ApJ...621..328H} and these methods have now been coupled to
a hydrogen chemical network \citep{2005ApJS..161...65A} and combined
with the 1893-level model of \Hmol{} described in
\citet{2005ApJ...624..794S}.

In this initial study, we restrict ourselves to models with elemental
abundances from \citet{1999ApJ...517..782H} that are illuminated by a
black-body source of luminosity 120~$L_\odot$ and effective
temperature 130,000~K, chosen to match the CSPN of the Helix nebula
(\citealp{1982ApJ...252..635B}, adjusted for the trigonometric
parallax of 217~pc, \citealp{2007AJ....133..631H}). We have also
calculated some models using a \citet{2003A&A...403..709R} stellar
atmosphere with the same luminosity and effective temperature.

We vary three different model parameters so as to span the range of
physical and illumination conditions that are seen in PN knots:
distance from the CSPN, $D$; hydrogen nuclei density at the
illuminated face, $n_0$; and gas velocity at the illuminated face,
$u_0$. Table~\ref{tab:input} summarises the input parameters of our
models.\footnote{Note that all flow velocities, $u$, are in the frame of
reference of the IF/DF, but since we find that $u$ becomes very small
at great depths, this is also approximately the rest frame of the
molecular gas.}

The magnitude of advective effects in the models is, to first order,
dependent only on the dimensionless combination: $\lambda_\mathrm{ad}
= n_0 u_0 4 \pi D^2 / Q_\mathrm{H}$, where $Q_\mathrm{H}$ is the
ionizing photon luminosity of the source. This \textit{advection
  parameter} \citep{2005ApJ...621..328H} is the ratio of particle flux
to photon flux, which increases with $u_0$ and decreases with
ionization parameter. The appropriate value of the downstream flow
velocity $u_0$ depends sensitively on the boundary conditions at the
illuminated face and on the global geometry of the flow. In the case
of a photoevaporating knot with negligible confining pressure on the
ionized side, $u_0$ will be of order the ionized sound speed ($\simeq
10~\kms$), and this is the case we concentrate on in this paper. In
the case of a more shell-like configuration of the molecular gas,
$u_0$ would tend to be lower.

\section{Predicted model structure}
\label{sec:pred-model-struct}

\begin{table*}
  \centering
  \setlength\tabcolsep{1.2\tabcolsep}
  \renewcommand\arraystretch{1.1}
  \caption{Physical conditions in different zones of a typical
    advective IF/DF structure}\label{tab:zones}
  \smallskip
  \vspace*{-\smallskipamount}
  \footnotesize
  \begin{tabular}{lllllllll}\hline
    Zone & Column (cm$^{-2}$)& $T$ (K)& $u$ (\kms)& $n_\mathrm{e}/n$ & $f_\Hmol$& $n/n_0$ & Heat & Cool \\ \hline
    I & $< 10^{18}$ & $10^4$ & $5$ & $0.6$ &$10^{-6}$ & 1 & H$^0$ p.e.& Metal, H$^0$ lines \\
    IIa & $1 \times 10^{18}$--$4 \times 10^{18} $ & 2000 & $0.5$ & 0.1 & 0.3 & 10 & H$^0$, He$^0$, H$_2$ p.e.& H$_2$ lines\\
    IIb & $4 \times 10^{18}$--$2 \times 10^{19}$ & 1000 & $0.2$ & 0.03 & 0.6 & 20 & He$^0$, H$_2$ p.e. &  H$_2$ lines\\
    III & $> 2 \times 10^{19} $ & 200 & $0.03$ & $3 \times 10^{-4}$ & 0.9 & $100$ & H$_2$ lines & FIR metal lines\\
    \hline
  \end{tabular}
\end{table*}

\begin{figure}
  \centering
  \includegraphics{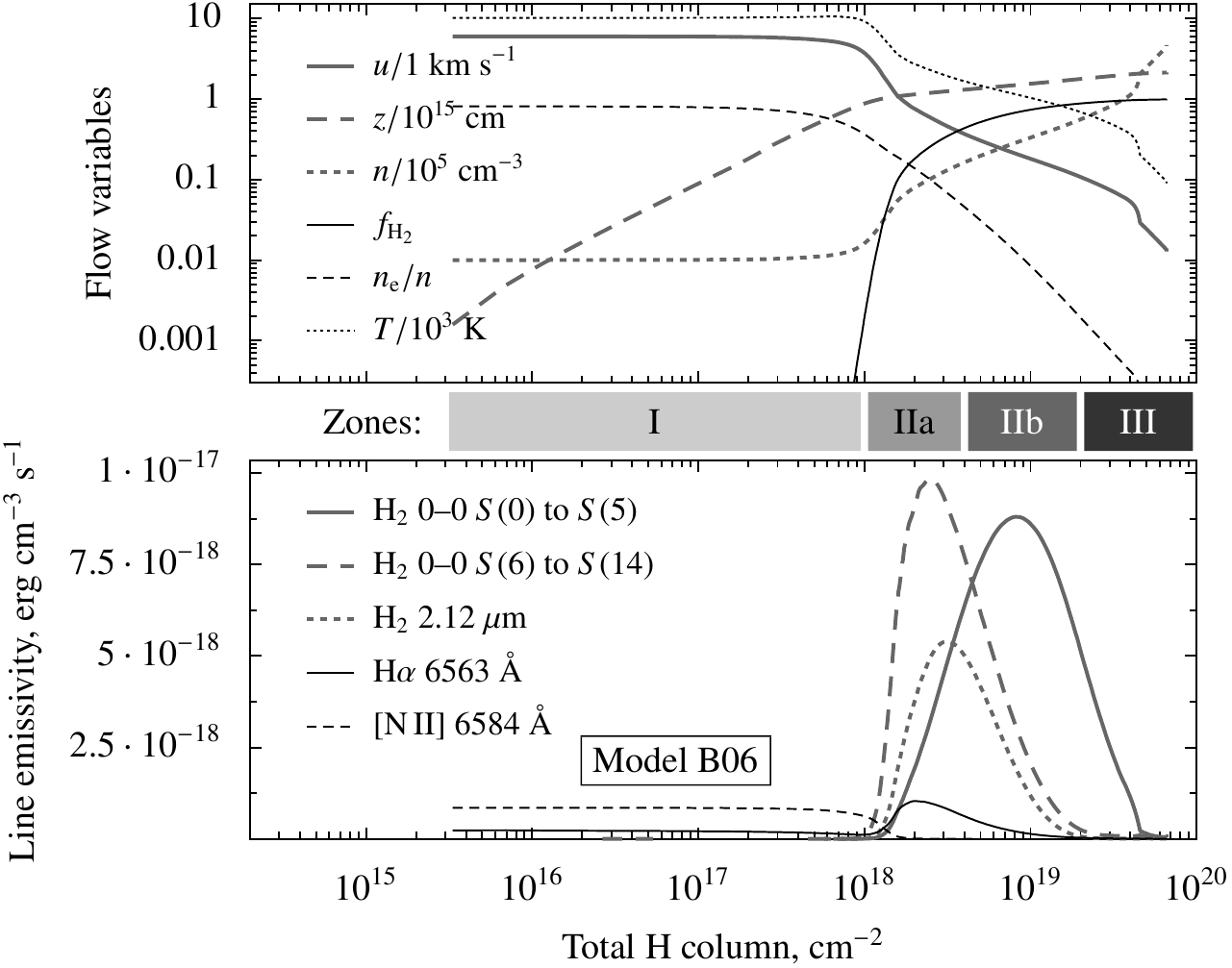}
  \caption{Structure of a representative advective model (B06, see
    Table~\protect\ref{tab:input}) as a function of column density of
    hydrogen nuclei. \textit{Top panel}: Various flow variables on a
    logarithmic scale: flow velocity, $u$, depth into cloud, $z$,
    total number density of hydrogen nuclei, $n$, hydrogen molecular
    fraction, $f_\Hmol$, electron fraction $n_\mathrm{e}/n$, and gas
    temperature, $T$.  \textit{Bottom panel}: Emissivity of important
    coolant lines.}
  \label{fig:flow-dynamic}
\end{figure}
\begin{figure}
  \centering
  \includegraphics{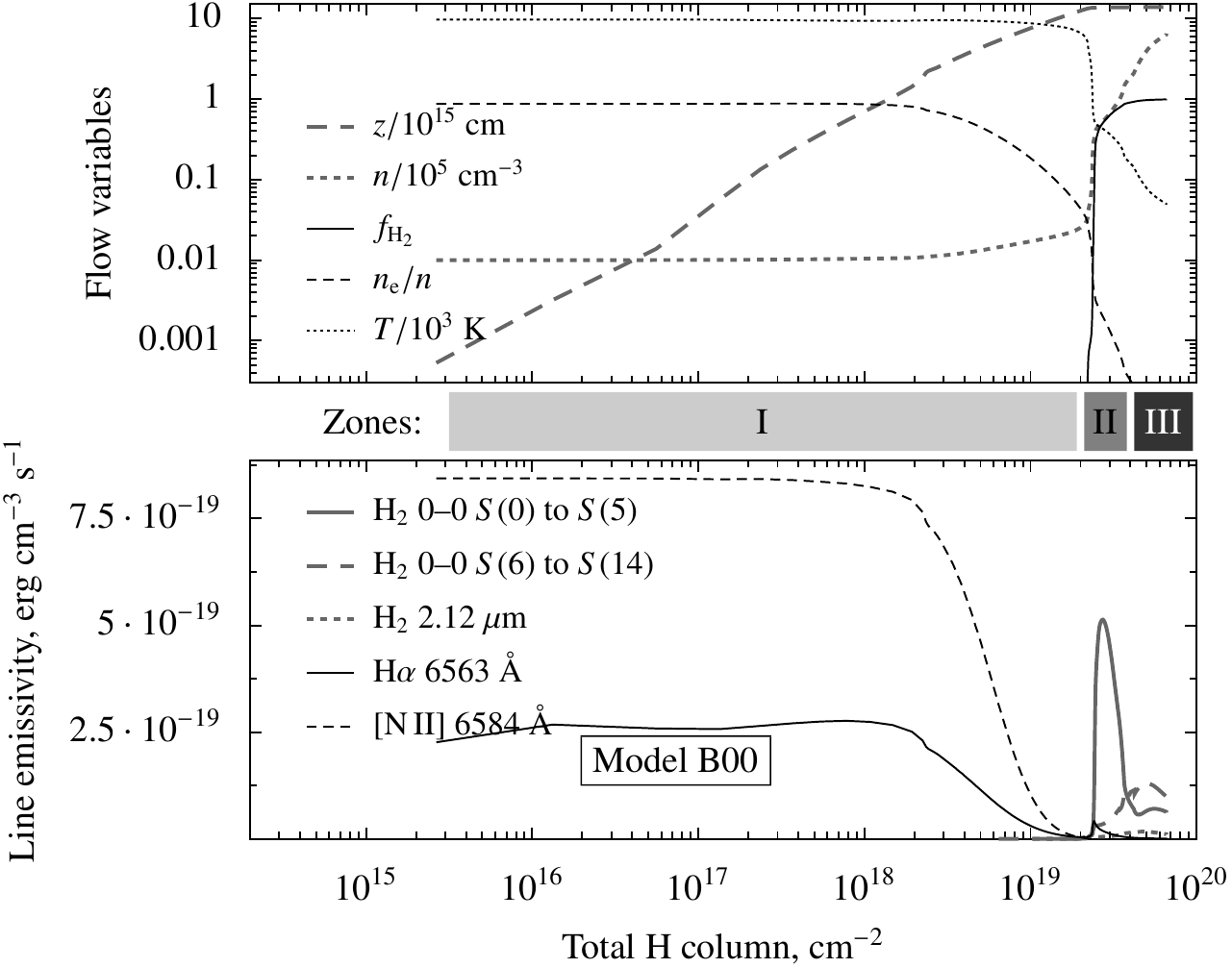}
  \caption{Same Fig.~\protect\ref{fig:flow-dynamic}, but for an
    equivalent static model (B00).}
  \label{fig:flow-static}
\end{figure}
\begin{figure}
  \centering
  \includegraphics{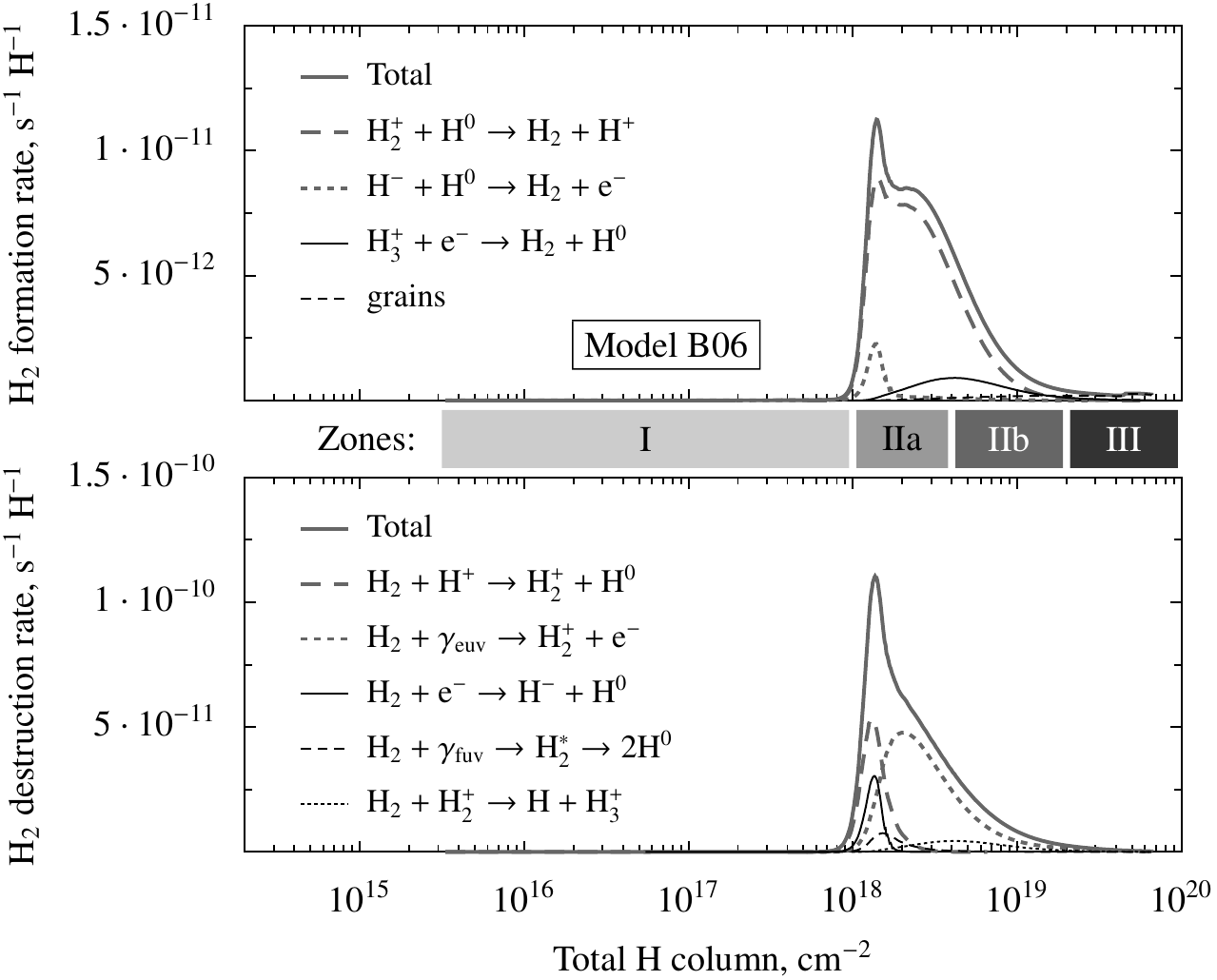}
  \caption{Destruction and formation rates of \Hmol{}, in units of
    molecules per hydrogen nucleus per second, for the advective model
    shown in Fig.~\protect\ref{fig:flow-dynamic}. \textit{Top panel}:
    Formation rates. \textit{Bottom panel}: Destruction rates. Note
    the different scales of the $y$ axes.  }
  \label{fig:dest-dynamic}
\end{figure}

\begin{figure}
  \centering
  \includegraphics{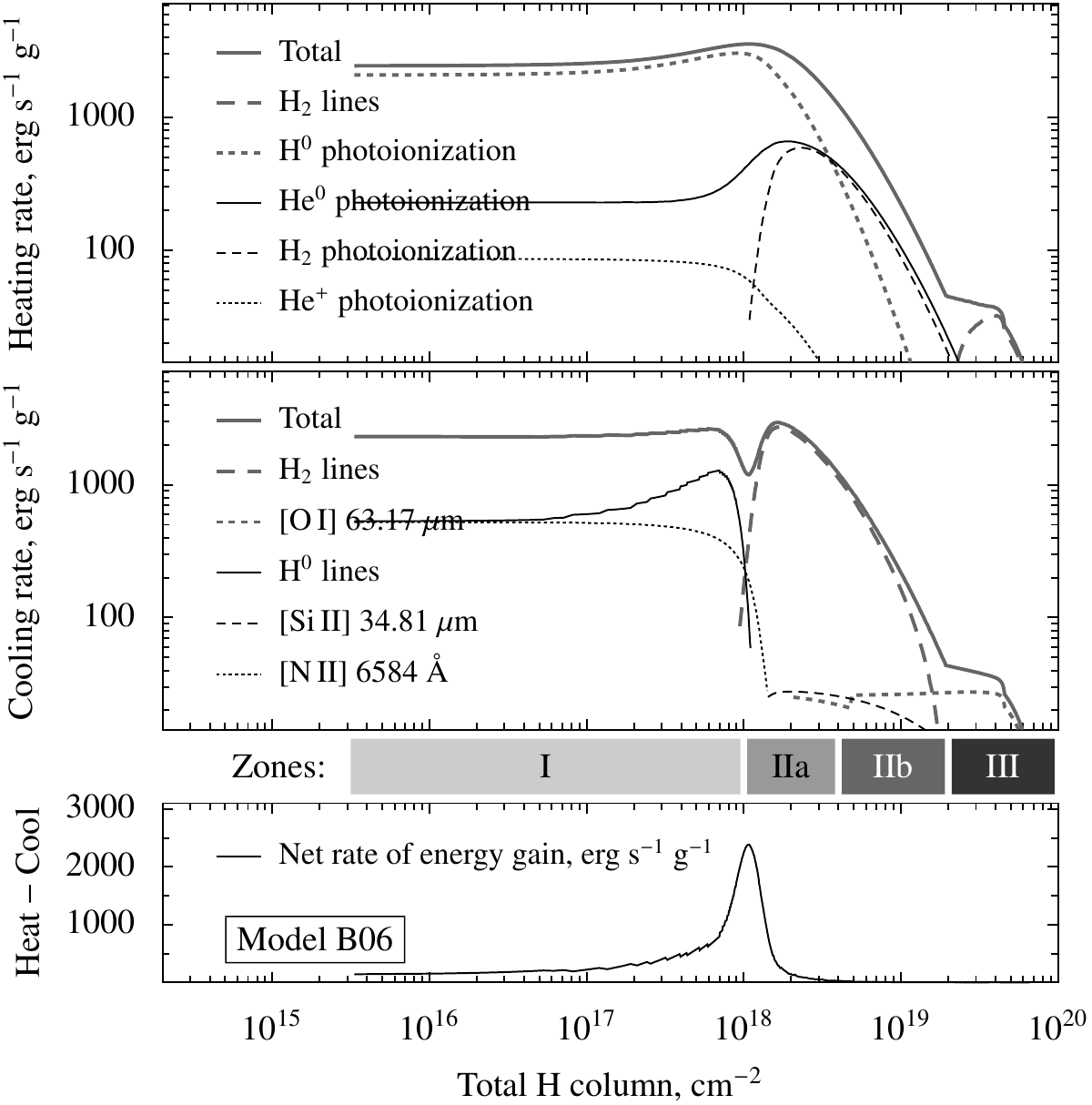}
  \caption{Thermal balance of the advective model shown in
    Fig.~\protect\ref{fig:flow-dynamic}. \textit{Top panel}: Principal
    contributions to the atomic/molecular heating rate per unit mass,
    shown on a logarithmic scale. \textit{Middle panel}: Same as top
    panel, but for cooling rate. \textit{Bottom panel}: Difference
    between heating and cooling rates, shown on a linear scale.}
  \label{fig:heatcool}
\end{figure}

Figure~\ref{fig:flow-dynamic} shows the results from a typical
advective model, B06, while Figure~\ref{fig:flow-static} shows results
from a static model, B00, with exactly the same incident flux and
density as B06.  For ease of discussion, we divide the model into 3
broad zones: Zone~I is closest to the ionizing source and is largely
ionized, with a very low molecular fraction; Zone~II is the
dissociation front, in which the hydrogen is half neutral and half
molecular (for the advective models, this zone is subdivided into IIa
and IIb); Zone~III is farthest from the ionizing source, where
hydrogen is fully molecular. Table~\ref{tab:zones} shows typical
physical conditions in each zone for the advective models. 

The differences between the advective and static models are stark: in
the advective model, the DF occurs at the very low column density of
$10^{18}\mathrm{\ cm^{-2}}$ from the illuminated face and strongly
overlaps with the ionization front, whereas in the static model the DF
occurs much deeper, at $2 \times 10^{19}\mathrm{\ cm^{-2}}$, in a
region where the ionization fraction is $< 0.01$. Zone~II is much
warmer in the advective model (2000~K in Zone~IIa, 1000~K in Zone~IIb)
than in the static model (500~K) and, as a result, \Hmol{} line
emission is more than an order of magnitude brighter. 

The rates of formation and destruction of \Hmol{} for Model~B06 are
shown in Figure~\ref{fig:dest-dynamic}, where it can be seen that the
destruction rate (bottom panel) has a narrow peak at the leading edge
of Zone~IIa, due to collisional processes with protons and electrons,
together with a broader peak covering Zones~IIa and IIb, due to
photoionization by hard EUV photons. The principal reaction channel
for the $\Hmol^+$ ions formed by these processes is dissociative
recombination with free electrons (e.g.,
\citealp{2007ApJ...659.1291M}), with only $\sim 10\%$ reacting with
H$^0$ to re-form \Hmol{}. Other \Hmol{} formation mechanisms have even
smaller rates (top panel), with the result that, once they have been
destroyed, most \Hmol{} molecules never re-form during the $\simeq
300$~yr that they remain in the DF.

The heating and cooling rates for Model~B06 are shown in
Figure~\ref{fig:heatcool}. In Zone~I, as is typical for low-excitation
\hii{} regions, the heating is dominated by H$^0$ photoelectric
heating, while the cooling is due to H lines and optical metal lines
such as [\ion{N}{2}] 6584~\AA{}. In Zone~IIa, H$^0$ photoelectric
heating still dominates the heating, whereas in Zone~IIb,
photoelectric heating of He$^0$ and \Hmol{} increasingly take over. In
all of Zone~II the cooling is overwhelmingly dominated by \Hmol{} line
emission. In Zone~III, the heating rate is much lower than in the
other zones and is due principally to the absorption of \Hmol{} lines
emitted in Zone~II, while the cooling in Zone~III is dominated by
collisionally excited far-infrared metal lines. The bottom panel of
Figure~\ref{fig:heatcool} shows the difference between the heating and
cooling rates, which is equal to the net rate of energy transfer from
the radiation field to the gas. This can be seen to have a sharp peak
at the boundary between Zones~I and IIa, where it represents a
significant fraction of the total heating. Outside this narrow heating
front, the gas is everywhere in approximate thermal equilibrium. The
fraction of the bolometric stellar luminosity that is converted into
thermal and kinetic energy of the gas can be shown to be $\simeq
\lambda_\mathrm{ad} T_0 / T_\mathrm{rad}$, where $T_0$ ($\simeq
10^4$~K) is the Zone~I gas temperature and $T_\mathrm{rad}$ ($\simeq
T_\mathrm{eff}$) is the color temperature of the incident
radiation. For the model shown, this fraction is 7\%, in good
agreement with the analytic estimate.

Other advective models show extremely similar structures. The shift in
column density of the DF with respect to the static models is roughly
proportional to $\lambda_\mathrm{ad}$, but even models with
$\lambda_\mathrm{ad} = 0.09$ have similar gas temperatures to
Model~B06 in Zones~IIa and IIb. Models using a
\citeauthor{2003A&A...403..709R} atmosphere instead of a black body
also produced extremely similar results, despite this spectrum having
a 20 times smaller flux at soft X-ray wavelengths ($>54.4$~eV).

A plane geometry is a poor approximation in Zone~I for the case of
photoevaporating knots, in which the ionized flow is expected to be
transonic and divergent \citep{2005AJ....130..172O}. However, the
small observed spatial offset between the \Hmol{}, H$\alpha$, and
[\ion{N}{2}] emission \citepalias{2007AJ....133.2343O} indicates that
the flow in Zones~II and III is approximately plane parallel. The flow
timescale through a column density, $N$, is equal to $N/n_0 u_0 \simeq
32 \lambda_\mathrm{ad} (N / 10^{18}~\mathrm{cm^{-2}})$~yr, which is
much less than the PN evolutionary timescale for the \Hmol{}-emitting
portions of the flow, justifying our steady-state assumption. However,
non-steady effects may be important in Zone~III, as may the formation
of CO and magnetic fields, neither of which are included in the
present models.

%

\begin{figure}
  \centering
  \includegraphics{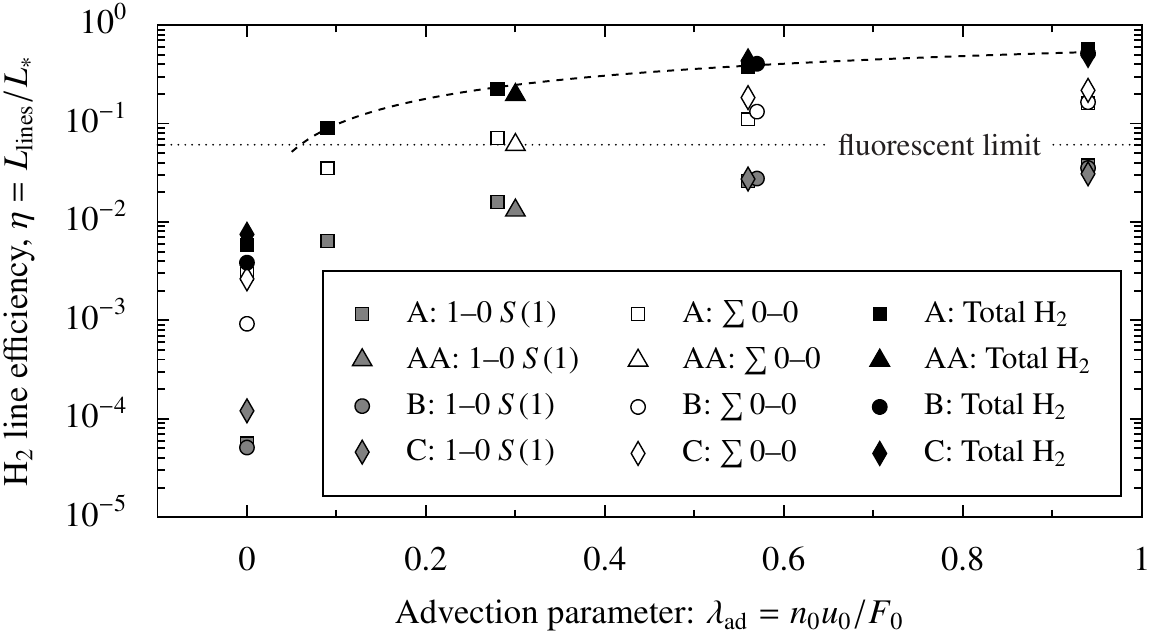}
  \caption{Fraction of the stellar bolometric luminosity that is
    radiated in \Hmol{} lines, assuming 100\% covering fraction of
    molecular gas. Black symbols show the total \Hmol{} line
    luminosity, white symbols show the sum of the pure rotational 0--0
    mid-infrared lines from $S(0)$ to $S(28)$, gray symbols show the
    near-infrared 1--0 $S(1)$ line at $2.121~\mu$m. Different symbol
    shapes are for different sequences of models, as shown in the
    key. The dashed curve shows an approximate analytic fit (see
    text).  }
  \label{fig:lumfrac}
\end{figure}

\begin{figure}
  \centering
  \includegraphics{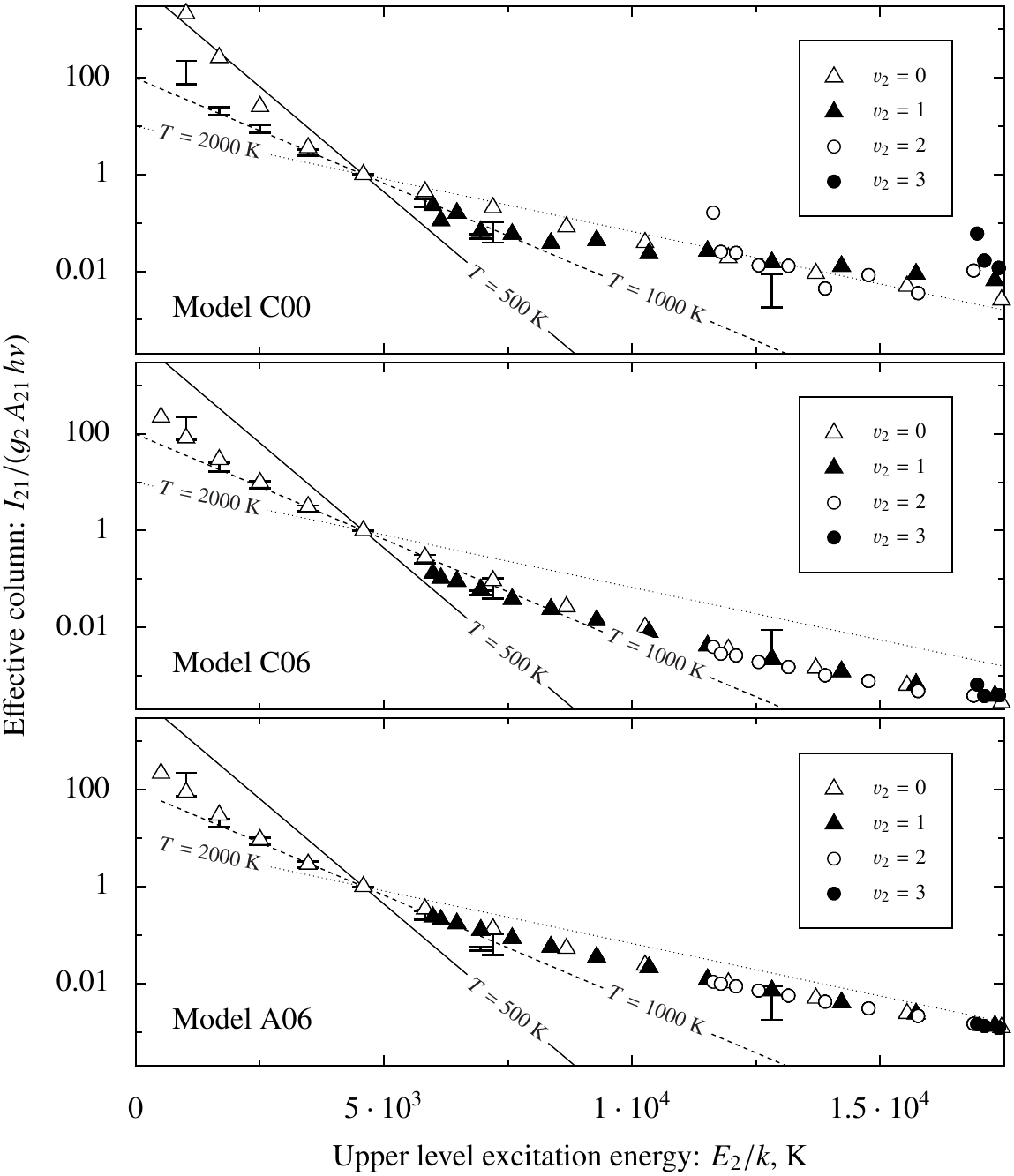}
  \caption{Excitation diagrams of \Hmol{} emission lines for three
    Cloudy models. \textit{Top:} Model~C00, \textit{Middle:}
    Model~C06, \textit{Bottom:} Model~A06. Symbols show the predicted
    effective columns from the Cloudy models, determined from the
    emergent line intensities, as a function of the excitation
    temperature of the upper level.  Different symbol types indicate
    the vibrational quantum number of the upper level, $v_2$, as shown
    in the key. The effective columns are normalised with respect to
    the value for the $v_2=0$, $J_2=7$ level, which gives rise to the
    0--0 $S(5)$ line at 6.907~$\mu$m.  Thin lines show the slope of
    Boltzmann distributions for 500~K (solid), 1000~K (dashed), and
    2000~K (dotted). Error bars show the range of measured values for
    the Helix knots (see text).}
  \label{fig:excit}
\end{figure}

\section{Predicted H$_2$ spectrum}
\label{sec:pred-h_2-spectr}
Figure~\ref{fig:lumfrac} shows the radiative efficiency of the models
in converting the stellar luminosity into \Hmol{} emission lines:
$\eta = L_\mathrm{lines} / L_*$. The value of $\eta_\mathrm{tot}$,
corresponding to the sum of all \Hmol{} lines is $< 0.01$ for the
static models, and rises rapidly with \lamad{} for the advective
models, approximately as $\eta_\mathrm{tot} = 1.1 \lamad / (1 +
\lamad^{0.9})$ (dashed curve in figure). The dotted line in the figure
indicates the maximum fluorescent efficiency, assuming that all
FUV radiation is converted into \Hmol{} lines.

Figure~\ref{fig:excit} shows predictions of the full \Hmol{} line
spectrum for transitions with upper level excitation temperatures $<
17,500$~K, which includes most near-infrared and mid-infrared
lines. Three representative models, are shown (\mbox{Table}~\ref{tab:input}):
a static model, C00, and two advective models, C06 and A06 with,
respectively, a low and a high incident flux. The line intensities are
shown in the standard way as effective column densities, which would
be proportional to $\exp(-E_2/k T)$ in the case of a Boltzmann
distribution at a single temperature, $T$, giving a straight line on
the semi-log plot.

All static models show a typical fluorescent spectrum with strong
deviations of the level populations from a Boltzmann distribution,
whereas advective models show a much smaller dispersion in the
effective columns of levels with similar excitation energies,
indicating that the excitation is largely thermal. The slope of the
excitation diagram is steeper for lower excitation energies, which can
be understood with reference Figure~\ref{fig:flow-dynamic}: the
emissivity of the lower pure rotational lines (excitation temperatures
$< 5000$~K) peaks in Zone~IIb, where the gas temperature is $\simeq
1000$~K, whereas higher excitation lines have their peak emissivity in
Zone~IIa, where the gas temperature is $\simeq 2000$~K.

\section{Discussion}
\label{sec:discussion}

Observations of the molecular hydrogen spectrum of the knots in the
Helix Nebula (\citealp{1998ApJ...495L..23C};
\citetalias{2006ApJ...652..426H}; \citetalias{2007AJ....133.2343O})
are indicated as error bars on Figure~\ref{fig:excit}. It can be seen
that the two advective models are in broad agreement with the
observations, whereas the static model is not. Model~C06 best matches
the distance of the spectroscopically observed knots from the ionizing
star, and indeed shows a better agreement than Model~A06, which
corresponds to the closer-in knots. The observed nebular flux in the
1--0 $S(1)$ line and in the sum of the 0--0 $S(1)$ to $S(7)$ lines are
$\simeq 1\%$ and $\simeq 4\%$, respectively, of the bolometric stellar
flux \citepalias{2007AJ....133.2343O}. From Figure~\ref{fig:lumfrac},
this can be satisfied by our models with $\lambda_\mathrm{ad} >
0.3/f$, where $f$ is the area covering fraction of knots. A strong
prediction of our model is that higher excitation lines from the $v
\ge 1$ levels should show a higher population temperature of $\simeq
2000$~K. A recent study of an inner knot \citep{2007arXiv0709.3065M}
finds a population temperature of about 1800~K for these levels, in
agreement with the prediction of our relevant model (A06).

Ro-vibrationally warm \Hmol{} has been detected in other PNe
\citep[e.g.,][]{2006AJ....131.1515L, 2007ApJ...659.1291M} and has
frequently been interpreted as evidence for shocks
\citep{1988ApJ...324..501Z}. However, EUV-dominated advective PDRs may
be a promising alternative in these cases too. 

\vspace*{-\baselineskip}
\acknowledgments We thank the following institutions and programs for
financial support: UNAM, Mexico (PAPIIT IN112006 and IN110108), STScI
(GO~10628 and AR~10653), NSF (AST~0607028), NASA (NNG05GD81G), and
Spitzer Science Center (20343).


\begin{thebibliography}{23}
\expandafter\ifx\csname natexlab\endcsname\relax\def\natexlab#1{#1}\fi
\expandafter\ifx\csname href\endcsname\relax
  \def\href#1#2{}\fi
\expandafter\ifx\csname urllinklabel\endcsname\relax
  \def\urllinklabel{[LINK]}\fi
\expandafter\ifx\csname adsurllinklabel\endcsname\relax
  \def\adsurllinklabel{[ADS]}\fi

\bibitem[{{Abel} {et~al.}(2005){Abel}, {Ferland}, {Shaw}, \& {van
  Hoof}}]{2005ApJS..161...65A}
{Abel}, N.~P., {Ferland}, G.~J., {Shaw}, G., \& {van Hoof}, P.~A.~M. 2005,
  \apjs, 161, 65


\bibitem[{{Abgrall} {et~al.}(2000){Abgrall}, {Roueff}, \&
  {Drira}}]{2000A&AS..141..297A}
{Abgrall}, H., {Roueff}, E., \& {Drira}, I. 2000, \aaps, 141, 297


\bibitem[{{Bertoldi} \& {Draine}(1996)}]{1996ApJ...458..222B}
{Bertoldi}, F. \& {Draine}, B.~T. 1996, \apj, 458, 222


\bibitem[{{Bohlin} {et~al.}(1982){Bohlin}, {Harrington}, \&
  {Stecher}}]{1982ApJ...252..635B}
{Bohlin}, R.~C., {Harrington}, J.~P., \& {Stecher}, T.~P. 1982, \apj, 252, 635


\bibitem[{{Cox} {et~al.}(1998){Cox}, {Boulanger}, {Huggins}, {Tielens},
  {Forveille}, {Bachiller}, {Cesarsky}, {Jones}, {Young}, {Roelfsema}, \&
  {Cernicharo}}]{1998ApJ...495L..23C}
{Cox}, P., 
et al.
1998, \apjl, 495, L23+


\bibitem[{{Ferland} {et~al.}(1998){Ferland}, {Korista}, {Verner}, {Ferguson},
  {Kingdon}, \& {Verner}}]{1998PASP..110..761F}
{Ferland}, G.~J., {Korista}, K.~T., {Verner}, D.~A., {Ferguson}, J.~W.,
  {Kingdon}, J.~B., \& {Verner}, E.~M. 1998, \pasp, 110, 761


\bibitem[{{Harris} {et~al.}(2007){Harris}, {Dahn}, {Canzian}, {Guetter},
  {Leggett}, {Levine}, {Luginbuhl}, {Monet}, {Monet}, {Pier}, {Stone},
  {Tilleman}, {Vrba}, \& {Walker}}]{2007AJ....133..631H}
{Harris}, H.~C., 
et~al.
2007, \aj, 133, 631


\bibitem[{{Henney} {et~al.}(2005){Henney}, {Arthur}, {Williams}, \&
  {Ferland}}]{2005ApJ...621..328H}
{Henney}, W.~J., {Arthur}, S.~J., {Williams}, R.~J.~R., \& {Ferland}, G.~J.
  2005, \apj, 621, 328


\bibitem[{{Henry} {et~al.}(1999){Henry}, {Kwitter}, \&
  {Dufour}}]{1999ApJ...517..782H}
{Henry}, R.~B.~C., {Kwitter}, K.~B., \& {Dufour}, R.~J. 1999, \apj, 517, 782


\bibitem[{{Hollenbach} \& {Tielens}(1997)}]{1997ARA&A..35..179H}
{Hollenbach}, D.~J. \& {Tielens}, A.~G.~G.~M. 1997, \araa, 35, 179


\bibitem[{{Hora} {et~al.}(2006){Hora}, {Latter}, {Smith}, \&
  {Marengo}}]{2006ApJ...652..426H}
{Hora}, J.~L., {Latter}, W.~B., {Smith}, H.~A., \& {Marengo}, M. 2006, \apj,
  652, 426 (H06)


\bibitem[{{Likkel} {et~al.}(2006){Likkel}, {Dinerstein}, {Lester}, {Kindt}, \&
  {Bartig}}]{2006AJ....131.1515L}
{Likkel}, L., {Dinerstein}, H.~L., {Lester}, D.~F., {Kindt}, A., \& {Bartig},
  K. 2006, \aj, 131, 1515


\bibitem[{{Matsuura} {et~al.}(2008){Matsuura}, {Speck}, {Smith}, {Zijlstra},
  {Viti}, {Lowe}, {Redman}, {Wareing}, \& {Lagadec}}]{2007arXiv0709.3065M}
{Matsuura}, M., 
et al.
2008,
  \mnras, in press, arXiv: 0709.3065


\bibitem[{{McCandliss} {et~al.}(2007){McCandliss}, {France}, {Lupu}, {Burgh},
  {Sembach}, {Kruk}, {Andersson}, \& {Feldman}}]{2007ApJ...659.1291M}
{McCandliss}, S.~R., 
et al.
2007, \apj, 659, 1291


\bibitem[{{Meixner} {et~al.}(2005){Meixner}, {McCullough}, {Hartman}, {Son}, \&
  {Speck}}]{2005AJ....130.1784M}
{Meixner}, M., {McCullough}, P., {Hartman}, J., {Son}, M., \& {Speck}, A. 2005,
  \aj, 130, 1784


\bibitem[{{Natta} \& {Hollenbach}(1998)}]{1998A&A...337..517N}
{Natta}, A. \& {Hollenbach}, D. 1998, \aap, 337, 517


\bibitem[{{O'Dell} {et~al.}(2005){O'Dell}, {Henney}, \&
  {Ferland}}]{2005AJ....130..172O}
{O'Dell}, C.~R., {Henney}, W.~J., \& {Ferland}, G.~J. 2005, \aj, 130, 172


\bibitem[{{O'Dell} {et~al.}(2007){O'Dell}, {Henney}, \&
  {Ferland}}]{2007AJ....133.2343O}
---. 2007, \aj, 133, 2343 (ODH07)


\bibitem[{{Rauch}(2003)}]{2003A&A...403..709R}
{Rauch}, T. 2003, \aap, 403, 709


\bibitem[{{Shaw} {et~al.}(2005){Shaw}, {Ferland}, {Abel}, {Stancil}, \& {van
  Hoof}}]{2005ApJ...624..794S}
{Shaw}, G., {Ferland}, G.~J., {Abel}, N.~P., {Stancil}, P.~C., \& {van Hoof},
  P.~A.~M. 2005, \apj, 624, 794


\bibitem[{{Stecher} \& {Williams}(1967)}]{1967ApJ...149L..29S}
{Stecher}, T.~P. \& {Williams}, D.~A. 1967, \apjl, 149, L29


\bibitem[{{Young} {et~al.}(1999){Young}, {Cox}, {Huggins}, {Forveille}, \&
  {Bachiller}}]{1999ApJ...522..387Y}
{Young}, K., {Cox}, P., {Huggins}, P.~J., {Forveille}, T., \& {Bachiller}, R.
  1999, \apj, 522, 387


\bibitem[{{Zuckerman} \& {Gatley}(1988)}]{1988ApJ...324..501Z}
{Zuckerman}, B. \& {Gatley}, I. 1988, \apj, 324, 501


\end{thebibliography}

\end{document}